\documentclass[10pt,a4paper]{book}
\usepackage{frontier04}
\begin{document}
\newcounter{ctr}
\setcounter{ctr}{\thepage}
\addtocounter{ctr}{8}

\talktitle{Blazing Cerenkov Flashes at the Horizons by  Cosmic
Rays and Neutrinos Induced Air-Showers}
\talkauthors{Daniele Fargion \structure{a,b}
             }
\begin{center}
\authorstucture[a]{Dipartimento di Fisica,
                   Universit\`a di Roma "La Sapienza",  \newline
                   Pl.A. Moro 2, Roma, Italy}
\authorstucture[b]{INFN, Sezione di Roma I, Italy}
\end{center}
\shorttitle{Blazing  at the Horizons}
\firstauthor{D. Fargion}

\begin{abstract}
High Energy Cosmic Rays (C.R.) versus  Neutrino
 and Neutralino induced Air-Shower maybe tested at  Horizons
by their muons, gamma and Cerenkov blazing signals. Inclined and
Horizontal C.R. Showers ($70^o-90^o$ zenith angle) produce
secondary ($\gamma$,$e ^\pm$) mostly suppressed
 by high column  atmosphere  depth. The air column depth  suppresses low energy Showers (TeV-PeV) and it dilutes higher
 energy  (PeV-EeVs) ones.
 Indeed also earliest  shower  Cherenkov  photons are diluted by large distances and by air opacity,
  while secondary penetrating, $\mu^\pm$  and their successive
  decay into $e^\pm$,$\gamma$, may revive additional Cerenkov lights. The
larger horizontal distances widen the shower's cone while the
geo-magnetic field open it in a  very characteristic fan-like
shape polarized by local field vector, making these elongated
showers spread, fork-shaped diluted and more frequent, up to three
order of magnitude (respect to vertical showers). GeVs $\gamma$
telescopes at the top of the mountains or in Space may detect at
horizons PeVs up to EeV and more energetic hadronic cosmic rays
secondaries. Details on arrival angle and column depth, shower
shape, timing signature of photon flash intensity, may inform us
on the altitude interaction and primary UHECR composition. Below
the horizons, at zenith angle ($90^o-99^o$) among copious single
albedo muons, rare up-going showers traced by muon
($e^\pm$,$\gamma$) bundles would give evidence of rare
Earth-Skimming neutrinos, ${\nu}_{\tau}$,
$\overline{\nu}_{\tau}$, at EeVs energies. They are arising by Tau
Air-Showers (HorTaus) (${\nu}_{\tau} + N \rightarrow \tau + X $,
$\tau\rightarrow $ hadrons and/or electromagnetic shower). Their
rate may be comparable with $6.3$ PeVs $\overline{\nu}_{e}-e$
neutrino induced air-shower (mostly hadronic) originated above
and also below horizons,  in interposed atmosphere by $W^-$
resonance at Glashow peak. Additional and complementary UHE SUSY
$\chi^o + e\rightarrow \tilde{e} \rightarrow\chi^o + e $ at tens
PeVs-EeV energy may blaze, as $\overline{\nu}_{e}-e \rightarrow
W^-$ shower, by its characteristic electromagnetic signature.
(These UHE $\chi^o $ are expected in topological defect scenarios
for UHECRs). Their secondary shower blazing Cerenkov lights and
distances  might be disentangled from UHECR by Stereoscopic
Telescopes such as Magic ones or Hess array experiment. The
horizontal detection sensitivity of Magic in the present set up
(if  totally devoted to the Horizons Shower search) maybe already
be comparable to AMANDA underground neutrino detector at PeVs
energies.

\end{abstract}
\section{UHECR Cerenkov Lights, Muons and Bundles at Horizons}
Ultra High Energy Cosmic Rays (UHECR) Showers (from PeVs up to
EeVs and above, mainly of hadronic nature) born at the high
altitude in the atmosphere, may blaze (from the far edge)
\emph{above  the horizon} toward Telescopes such as Magic one. The
earliest gamma and Cerenkov lights produced while they propagate
through the atmosphere are severely absorbed because of the deep
horizontal atmosphere column depth ($10^4 - 5 \cdot 10^4$ $ g
\cdot cm^{-2}$ ),
  must anyway survive and also revive: indeed additional diluted but  penetrating
   muon bundles (from the same by C.R. shower)
   are decaying not far from the Telescope into electrons which are source themselves of small Cerenkov lights .
  Also  the same muon while hitting the Telescope may blaze a ring
  or arc of Cherenkov lights.  These suppressed muon bundle secondaries,( about $10^{-3}$ times
  less abundant than the  peak of the gamma shower photons) are  arising at high altitude, at an horizontal
  distances  $100-500$ km far from the observer (for a zenith angle
  $85^o-91.5^o$ while at 2.2km. height);
  therefore these hard (tens-hundred GeV) muon shower bundles (from  ten to millions muons from TeVs-EeVs C.R. primary)
   might spread in huge areas (tens- hundred $km^2$); they are   partially bent by geo-magnetic fields
    and they are randomly scattered, often decaying at tens-hundred GeV energies,
     into electrons and consequent mini electromagnetic-showers traced by their optical Cerenkov flashes.
  These diluted (but spread and therefore better detectable) brief (nanosecond-microsecond)
   optical signals may be captured as a cluster by largest  telescope on ground
    as recent Stereoscopic Magic or Hess, Veritas arrays.
    Their  Cerenkov flashes, single or  clustered,
   must take place, at detection threshold, at least tens or hundreds times a night
   for Magic-like Telescope facing toward horizons $85^o-90^o$.
   Their "guaranteed" discover may offer  a new tool in  CR and UHECR detection.
   Their primary hadronic signature might be hidden by the distance but its tail may arise in a new form by
   its secondary muon-electron-Cerenkov  of electromagnetic
   nature. On the same time  \emph{below the horizons}  a more rare (three-four order of
 magnitude)  but  more exciting PeV-EeVs Neutrino ${\nu_{\tau}}$
   Astronomy may arise by the Earth-Skimming Horizontal Tau Air-Showers (HorTaus); these UHE Taus are produced
   inside the Earth Crust by the primary UHE incoming neutrino
      ${\nu}_{\tau}$, $\overline{\nu}_{\tau}$,
       generated mainly by their muon-tau neutrino oscillations from
       galactic or cosmic sources,\cite{Fargion 2002a},\cite{Fargion03},\cite{Fargion2004}.
   Finally just  above or below the horizon edge, within a
   few hundred of km distances,  it might also be observable the
   guaranteed and well tuned   $\overline{\nu}_e$-$e\rightarrow W^-\rightarrow X$ air-showers at $6.3 PeV$ Glashow resonant peak
    energy; the W main  hadronic ($2/3$) or leptonic and electromagnetic ($1/3$) signatures  may be well observed and
     their rate might calibrate a new horizontal neutrino-multi-flavour
     Astronomy \cite{Fargion 2002a}. The  $\overline{\nu}_e$-$e\rightarrow W^-\rightarrow X$
     of nearby nature (respect to most far away ones at same zenith angle of hadronic nature) would be better revealed by
     a Stereoscopic Magic  twin telescope or a Telescope array like Hess, Veritas.
       Additional Horizontal flashes  might arise
    by Cosmic UHE $\chi_o + e \rightarrow  \widetilde{e}\rightarrow \chi_o +
    e$ electromagnetic showers  within most SUSY models, if UHECR are born in topological
   defect decay or in their annihilation, containing a relevant component of SUSY
   particles. The UHE $\chi_o + e \rightarrow  \widetilde{e}\rightarrow \chi_o +
    e$ behaves (for light $\widetilde{e}$ masses around Z boson ones)
    as the Glashow  resonant case \cite{Datta}.
    Finally similar signals might be abundantly and better observed if UHE neutrinos share new extra-dimension
    (TeV gravity) interactions: in this case also  neutrino-nucleons interaction may be an abundant source of PeVs-EeVs
    Horizontal Showers originated in Air \cite{Fargion 2002a}. The total amount of air
    inspected within the solid angle $2^o \cdot 2^o$ by MAGIC height  at  Horizons ($360$ km.) exceed
    $44 km^3$ but their consequent detectable beamed volume  are  corresponding
    to an isotropic  narrower volume:  V$ = 1.36 \cdot 10^{-2}$ $km^3$ , nevertheless comparable
     (for Pevs $\overline{\nu}_e$-$e\rightarrow W^-\rightarrow X$
      and EeVs  ${\nu}_{\tau}$, $\overline{\nu}_{\tau} + N \rightarrow \tau\rightarrow $ showers) to the  present AMANDA confident volume.

\begin{figure}
\begin{center}
\epsfig{figure=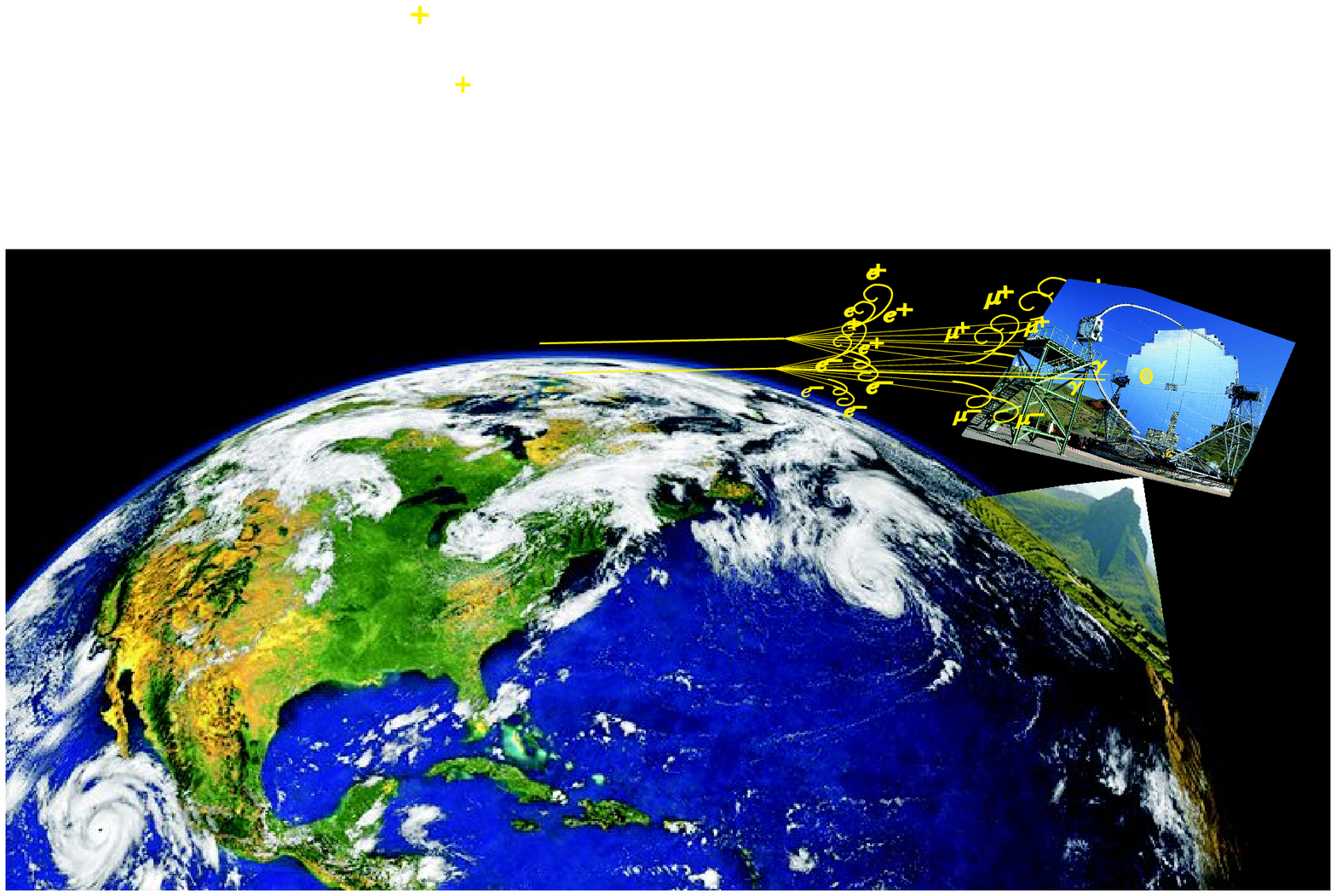,width=8cm}

\end{center}
\caption{ Schematic Picture of an Horizontal Cosmic Ray Air-Shower
(superior track) (HAS),  and an up-going Tau Air-Shower induced
by EeV Earth-Skimming $\bar{\nu_{\tau}}$,$\nu_{\tau}$ HORTAU and
their muons and Cerenkov lights blazing a Telescope as the Magic
one. Also UHE $\bar{\nu_e}-e$ and $\chi^o - e$ Scattering in
terrestrial horizontal atmosphere at tens PeVs energy may
simulate HAS, but mostly at nearer distances respect largest EeV
ones of hadron nature at horizon's edges. }

\begin{center}
\epsfig{figure=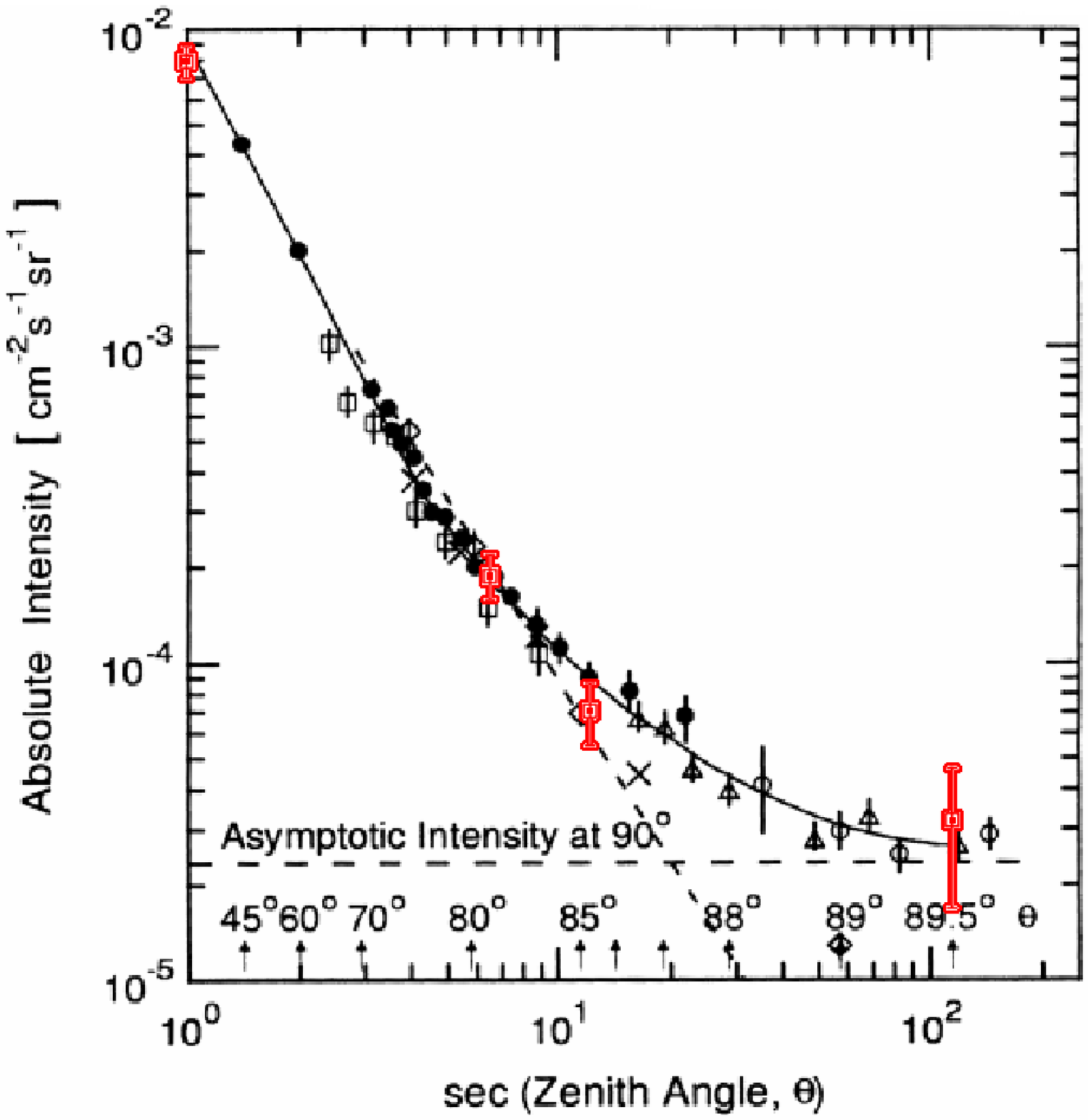,width=8cm}
\end{center}
\caption{ Observed Flux of Muons as a function of the zenith angle
above (left \cite{Iori04},\cite{Grieder01}) the horizons; for the
muons below the horizons their flux at $91^o$ zenith angle is two
order of magnitude below $\simeq 10^{-7} cm^{-2} s^{-1} sr^{-1}$,
as observed by NEVOD and Decor detectors in recent years. As the
zenith angle increases the upward muons flux reduces further ; at
$94^o$ and ten GeV energy it is just four order below: $\simeq
10^{-9} cm^{-2} s^{-1} sr^{-1}$ \cite{NEVOD},\cite{Decor}; at
higher energies (hundred GeVs) and larger zenith angle only muons
induced by atmospheric neutrinos arises at $\simeq 2-3\cdot
10^{-13} cm^{-2} s^{-1} sr^{-1}$ as well as Neutrino Tau induced
Air-Shower (muon secondaries).
 }
\end{figure}

\section{ Blazing Cerenkov Flashes by Horizons Showers and Muons }
  The ultrahigh energy cosmic rays (UHECR) have been studied
   in the past mainly versus their secondaries ($\gamma$, $e^\pm$, $\mu^\pm$)
  collected vertically in large  array detectors on the ground. This is due to the rare event rate
  of the UHECR  in the atmosphere and due to the high
  altitude where the shower takes place, expand and amplify downward.
  On the contrary at the horizons the UHECR  are hardly observable (but also rarely looked
  for).  They are diluted both by  the larger  distances as well as by the exponential
  atmosphere opacity suppressing the electromagnetic (electron pairs and gamma) secondaries;
   also their rich optical Cerenkov signal is partially  suppressed by the horizontal air opacity.
   However this suppression acts also as a useful filter
  leading to the choose of higher CR events; their Cerenkov lights
  will be scatter and partially transmitted ($1.8\cdot 10^{-2}$ at $551$ nm., $6.6\cdot 10^{-4}$ at $445$ nm.)
  depending on the exact zenith angle and seeing: assuming a suppression on average $5\cdot 10^{-3}$
  and  the  nominal Magic threshold at $30$ GeV  gamma energy, it  does
    corresponds to a hadronic shower at far horizons (diluted by nearly three order of magnitude by larger distances)
     at an  energy above $E_{CR}\simeq 6 $ PeV.  Their primary flux maybe estimated considering
     the known cosmic ray fluxes at same energy on the top of the atmosphere (both protons
    or helium) (see DICE Experiment referred in\cite{Grieder01}) :
     $\phi_{CR}(E_{CR} = 6\cdot 10^{15} eV)\simeq 9\cdot 10^{-12}cm^{-2}s^{-1}$;
    the consequent event rate spread within a Shower Cerenkov angle $\Delta\theta = 1^o $
     at a distance  $d =167 km \cdot \sqrt{\frac{h}{2.2 km}}$
     (zenith angle $\theta \simeq 87^o- 88^o$)
     corresponding to a wide shower area $ [A = \pi \cdot(\Delta\theta \cdot d)^2\simeq 2.7 \cdot 10^{11} cm^2 (\frac{d}{167 km})^2 ]$, observed
     by a opening angle $[\Delta\Omega =(2^o \cdot 2^o)\pi \simeq 3.82  \cdot 10^{-3} sr.]$
     is  for a night of record ($[\Delta(t)= 4.32 \cdot 10^4 s]$):
      $$N_{ev}=\phi_{CR}(E= 6\cdot 10^{15} eV)\cdot A \cdot \Delta \Omega \cdot
      \Delta(t) \simeq 401/12 h$$ Therefore one may foresee  nearly every
      two minutes  a far hadronic Cerenkov lightening  Shower in Magic facing at the far horizons
      at zenith angle $87^o-88^o$.  Increasing the observer altitude h, the
      horizon zenith angle also grows: $\theta \simeq [90^o + 1.5^o \sqrt{\frac{h}{2.2km}}]$
       In analogy at a more distant horizontal edges (standing at height $2.2
       km$ as for Magic, while observing at zenith angle $\theta \simeq 89^o- 91^o$
         still above the horizons) the observation range $d$ increases : $d= 167\sqrt{\frac{h}{2.2 km}} + 360 km = 527
       km$;  the consequent shower area widen by more than an order of
       magnitude (and more than  three order respect to vertical showers) and the foreseen event number,
       now for much harder C.R. at $E_{CR} \geq 3\cdot 10^{17} eV$,  becomes:
        $$N_{ev}=\phi_{CR}(E= 3\cdot 10^{17} eV)\cdot A \cdot \Delta \Omega \cdot
      \Delta(t) \simeq 1.6 /12 h$$
       Therefore at the far edges of  the horizons $\theta \simeq 91.5^o$, once a night, an UHECR around EeV energies,
      blazes to the Magic (or Hess,Veritas, telescopes).
        At each of these far primary Cherenkov flash is associated a
        long trail of secondary muons in a very huge area; these muons eventually are also hitting inside the
        Telescope disk; their nearby showering, while decaying
        into electrons in flight, (source of  tens-hundred GeVs  mini-gamma showers)
        is  also detectable at a rate discussed below.

\section{Single-Multi muons: Arcs, Rings and Gamma by
$\mu^\pm \rightarrow \gamma, e^\pm$}

    As already noted the main shower blazing photons from a CR  may be also reborn or overlap with its secondary
    tens-hundred GeVs muons, either  decaying in flight as a gamma
    flashes,  or by direct Cerenkov muons lights painting arcs or rings while hitting the telescope.
    Indeed these secondary very penetrating muon bundles
    may reach hundreds km far distances ($\simeq 600 km \cdot\frac{E_{\mu}}{100\cdot GeV}$) away from the shower origin.
    To be more precise a part of the muon primary energy will dissipate along $360$ km air-flight (nearly a
      hundred GeV energy), but a primary $130-150$ GeV. muon will reach a final
       $30-50$ GeV energy, just at minimal  Magic threshold value.
    Let us remind the characteristic secondary abundance in a shower:
    $ N_\mu \simeq 3\cdot 10^5 \left( \frac{E_{CR}}{PeV}\right)^{0.85} $
     These multiplicity are just at a minimal (GeV) energies
\cite{Cronin2004};  for the harder (a hundred
   GeV) muons their number is (almost inversely proportionally to energy) reduced:
  $ N_\mu(10^2\cdot GeV) \simeq 1.3\cdot 10^4 \left( \frac{E_{CR}}{6 \cdot PeV}\right)^{0.85}$
   These values must be compared with the larger peak multiplicity (but much lower energy) of
   electro-magnetic shower  nature: $ N_{e^+ e^-} \simeq 2\cdot 10^7 \left(
   \frac{E_{CR}}{PeV}\right); N_{\gamma} \simeq  10^8 \left(
   \frac{E_{CR}}{PeV}\right) $.  As mentioned most of these electromagnetic tail  is lost (exponentially) at
  horizons (above slant depth of a few hundreds of
  $\frac{g}{cm^2}$)(out of the case of re-born, upgoing $\tau$ air-showers
  \cite{Fargion2004},\cite{Fargion2004b}); therefore
   gamma-electron pairs are only partially  regenerated
    by the penetrating muon decay in flight, $\mu^\pm \rightarrow \gamma, e^\pm$
   as a parasite  electromagnetic showering \cite{Cillis2001}.
   Indeed $\mu^\pm $  may decay in flight (let say  at $100$ GeV energy,at $2-3\%$ level within a $12-18$ km distances)
    and they may inject more and more lights, to their primary (far born) shower beam.

   These tens-hundred GeVs  horizontal muons and their associated mini-Cerenkov $\gamma$ Showers have two  main origin:
    (1) either a single muon mostly produced at hundreds km distance by a single (hundreds GeV-TeV
       parental) C.R. hadron primary (a very dominant component)
    or
    (2) rarer muon, part of a wider and spread horizontal muon  bundle
    of large multiplicity born at TeVs-PeV  or higher energies, as secondary of horizontal shower.
     Between the two cases there is a smooth link.
     A whole continuous spectrum  of multiplicity  begins from an unique muon up to a multi muon shower production.
     The  dominant noisy "single" muons at hundred-GeV energies
     will loose memory of their primary low energy and  hidden  mini-shower, (a hundreds GeV or TeVs hadrons );
      a single muon  will blaze just alone.
    The muon "single" rings or arcs frequency is larger (than muon bundles ones) and it is based on solid observational data
    (\cite{Iori04} ; \cite{Grieder01},as shown in fig.2  and references on MUTRON experiment therein); these "noise" event number is:
     $$N_{ev}= \phi_{\mu}(E\simeq 10^{2} eV) \cdot A_{Magic} \cdot \Delta \Omega \cdot
      \Delta(t) \simeq 120 /12 h$$
      The additional gamma  mini-showers around the telescope due to a decay
        (at a probability $p\simeq 0.02$) of those muons in flight, recorded within a
         larger collecting  Area $A_{\gamma} \geq 10^9 cm^2$ is even a more frequent (by a factor $\geq 8$) noisy signal:
       $$N_{ev}\geq \phi_{\mu}(E\simeq 10^{2} eV)\cdot p \cdot A_{\gamma} \cdot \Delta \Omega \cdot
      \Delta(t) \simeq 960 /12 h$$  These   single background gamma-showers must take place nearly once
       a minute (in an silent hadronic background) and they are an useful  tool
       to be used as a prompt meter of the Horizontal C.R. verification.



    On the contrary PeVs (or higher energy) CR shower Cerenkov lights
     maybe  observed, more rarely, in coincidence  both by their primary and by their later secondary arc and gamma mini-shower.
   Their $30-100$ GeV  energetic muons are flying  nearly undeflected
  $\Delta \theta \leq 1.6^o \cdot \frac{100 \cdot GeV}{E_{\mu}}\frac{d}{300 km}$
  for a characteristic horizons distances d , partially bent by  geo-magnetic $0.3$  Gauss fields;
  as mentioned, to flight   through the whole horizontal air column depth
  ($360$ km equivalent to $360$ water depth) the muon
   lose nearly $100$ GeV; consequently the origination muon energy should be a little  above this threshold
   to be observed by Magic: (at least $ 130-150 $ GeV along most of the flights).
   The deflection angle is therefore a small one:
    $\Delta \theta \leq 1^o \cdot \frac{150 \cdot GeV}{E_{\mu}}\frac{d}{300
   km}$). Magic telescope area ($A = 2.5 \cdot 10^6 cm^2$) may record at first approximation the
   following event number of  direct muon hitting the Telescope, flashing  as rings and arcs, each night:
  $$N_{ev}=\phi_{CR}(E= 6\cdot 10^{15} eV)\cdot N_\mu(10^2\cdot GeV) \cdot A_{Magic} \cdot \Delta \Omega \cdot
      \Delta(t) \simeq 45 /12 h$$ to be correlated (at $11\%$ probability) with the above
      results of $401$ primary Cerenkov flashes at the far distances.
   As already mentioned before, in addition the same muons are decaying in flight  at a minimal probability $2\%$
   leading to a  mini-gamma-shower event number in a quite wider  area ($A_{\gamma}= 10^9 cm$):
   $$N_{ev}= \phi_{CR}(E= 6\cdot 10^{15} eV)\cdot N_\mu(10^2\cdot GeV) \cdot p \cdot A_{\gamma} \cdot \Delta \Omega \cdot
      \Delta(t) \simeq 360 /12 h$$
      Therefore , in conclusion,  at $87^o-88^o$ zenith angle, there are a flow of
      primary $ E_{C.R}\simeq 6\cdot 10^{16} eV$ C.R. whose earliest showers and consequent secondary muon-arcs as well as
      nearby muon-electron mini-shower take place at comparable (one every
      $120$ s.) rate. These $certain$  clustered signals offer an unique tool for
      gauging and calibrating Magic (as well as
      Hess,Cangaroo,Veritas Cerenkov Telescope Arrays) for Horizontal High Energy Cosmic Ray Showers.
      Some more rare event may contain at once both Rings,Arcs and tail
      of gamma  shower and Cerenkov of far primary shower.
       It is possible to estimate also the observable muons-electron-Cerenkov  photons
  from up-going  Albedo muons observed by recent ground experiments  \cite{NEVOD}  \cite{Decor}: their flux
   is already suppressed at zenith angle $91^o$ by at least two order of
   magnitude and by four order for up-going zenith  angles $94^o$.
   Pairs or bundles are nevertheless more rare (up to $\phi_{\mu} \leq 3 \cdot 10^{-13} cm^{-2}s^{-1}sr^{-1}$
   \cite{NEVOD}  \cite{Decor}). They are never associated to up-going shower out of the case of
   tau air-showers or by nearby Glashow $\bar{\nu_e}-e\rightarrow W^-$ and  comparable $\chi^o + e\rightarrow \tilde{e}$
   detectable by stereoscopic Magic or Hess array telescopes, selecting and evaluating their column depth origination, just discussed below.

\section{UHE $\bar{\nu_e}-e\rightarrow W^-$ and $\chi^o + e\rightarrow \tilde{e}$ resonances versus $\tau$ air-showers }
  The appearance of horizontal UHE
   $\bar{\nu_\tau}$ ${\nu_\tau}\rightarrow \tau$ air-showers (Hortaus or Earth-Skimming neutrinos)
    has been widely studied  \cite{Fargion1999},\cite{Fargion 2002a},
  \cite{Bertou2002},\cite{Feng2002},\cite{Fargion03},\cite{Fargion2004},\cite{Jones04},
  \cite{Yoshida2004},\cite{Tseng03},\cite{Fargion2004b};  their rise from the Earth is source of rare clear signals for neutrino UHE astronomy (see fig.3).
   However also horizontal  events by UHE $6.3$ PeV, Glashow $\bar{\nu_e}-e\rightarrow W^-$ and a possible
   comparable SUSY  $\chi^o + e\rightarrow \tilde{e}$ \cite{Datta} hitting and showering in air have  non negligible event number:
   $$N_{ev}= \phi_{\bar{\nu_e}} (E= 6\cdot 10^{15} eV) \cdot A\cdot \Delta \Omega \cdot
      \Delta(t) \simeq  5.2 \cdot 10^{-4}/12 h$$
      assuming the minimal  GZK neutrino flux : $\phi_{\bar{\nu_e}} (E= 6\cdot 10^{15} eV)\simeq 5 \cdot
      10^{-15}$ eV $ cm^{-2} s^{-1}sr^{-1}$. Therefore
   during a year of night records and such a minimal GZK flux,  a crown array of
   a $90$ Magic-like telescopes on $2\cdot \pi = 360^o$ circle facing the horizons,
   would discover an event number  comparable to a
   $Km^3$ detector, ( nearly a dozen events a year). Indeed  Magic facing
   at the Horizons as it is, offer a detection comparable to
   present AMANDA $\simeq  1\%  Km^3$ effective volume. In conclusion while Magic looking
   up see Gamma GeV Astronomy, Magic looking at Horizons may well see
   UHE (PeVs-EeVs) CR, and rarely along the edge, GZK $\bar{\nu_e}-e\rightarrow W^-$ neutrinos, $\nu{\tau} \rightarrow \tau$ air-showers and, surprisingly
   even  SUSY $\tilde{e}$ lights in the sky (with showers).

\begin{figure}
\begin{center}
\epsfig{figure=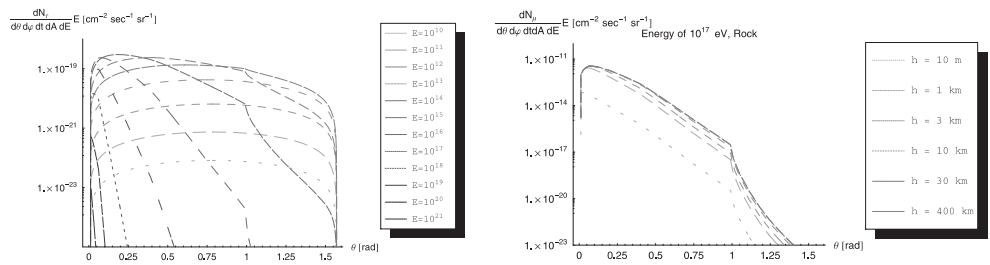,width=13cm} \caption{ Tau
Air-Showers (left) rates, by Earth Skimming Neutrino $\tau$ and
their consequent (right) Muons Secondary rate angular distribution
at different observer height, at $10^{17}$eV energy, exceeding
even the atmospheric neutrino induced muon  flux $\phi_{\mu}
\simeq 3 \cdot  10^{-13} \cdot cm^{-2} s^{-1}sr^{-1}$ }
   \label{fig3}
   \end{center}
\end{figure}

\end{document}